\documentclass{PoS}

\usepackage{xspace}

\newcommand{\etjet}{\ensuremath{E_{T}^{\rm{jet}}}\xspace}
\newcommand{\etajet}{\ensuremath{\eta^{\rm{jet}}}\xspace}
\newcommand{\alphas}{\ensuremath{\alpha_{s}}\xspace}
\newcommand{\alphasmz}{\ensuremath{\alpha_{s}(M_{Z})}\xspace}

\title{Inclusive Jets in PHP}

\ShortTitle{Inclusive jets in photoproduction}

\author{\speaker{Philipp Roloff}\thanks{On behalf of the ZEUS collaboration.}\\
        CERN\\
        E-mail: \email{philipp.roloff@cern.ch}}

\abstract{Differential inclusive-jet cross sections have been measured 
in photoproduction for boson virtualities $Q^{2} < 1$~GeV$^{2}$ with the ZEUS 
detector at HERA using an integrated luminosity of $300$~pb$^{-1}$. 
Jets were identified in the laboratory frame using the $k_{T}$, anti-$k_{T}$ or SIScone jet algorithms.
Cross sections are presented as functions of the jet pseudorapidity, \etajet, 
and the jet transverse energy, \etjet. Next-to-leading-order QCD calculations give a 
good description of the measurements, except for jets with low \etjet and high \etajet. 
The cross sections have the potential to improve the determination of the PDFs in future QCD fits. 
Values of \alphasmz have been extracted from the measurements based on different jet 
algorithms. In addition, the energy-scale dependence of the strong coupling was determined.}

\FullConference{The European Physical Society Conference on High Energy Physics \\
		18-24 July 2013\\
		Stockholm, Sweden}

\begin{document}

\section{Introduction}
\label{sec:introduction}

The measurement of jet photoproduction (PHP) at HERA provides a high-statistics 
test of perturbative QCD (pQCD) in a process with a single hard scale, \etjet. Jet 
cross sections allow precise determinations of the strong coupling constant, \alphas, and 
its energy dependence.

At leading order, so-called direct and resolved processes contribute to jet photoproduction. 
In direct processes, the photon interacts directly with a parton 
in the proton. On the other hand, the photon acts as a source of partons for the resolved 
contributions. Hence inclusive-jet cross sections are directly sensitive to the 
proton and photon PDFs.

The $k_{T}$ cluster algorithm~\cite{ref:kt_algorithm} in the longitudinally invariant inclusive 
mode~\cite{ref:long_inv_incl_mode} results in small theoretical uncertainties and hadronisation corrections in electron-proton 
collisions. It yields infrared- and collinear-safe cross sections at any order of QCD. More recently, new 
infrared- and collinear-safe algorithms like anti-$k_{T}$~\cite{ref:anti_kt_algorithm} 
or SIScone~\cite{ref:siscone_algorithm} were developed. Jet photoproduction at HERA represents 
a well-understood hadron-induced reaction to test and compare the performances of these different 
jet clustering algorithms.

NLO QCD predictions~\cite{ref:nlo_predictions} are compared to the measured inclusive-jet cross 
sections. The renormalisation and factorisation scales were set to $\mu_{R} = \mu_{F} = \etjet$ and the 
number of flavours was chosen to be five. Unless explicitly stated otherwise, the ZEUS-S~\cite{ref:zeus_s_pdfs} 
parametrisations were used for the proton PDFs and the GRV-HO~\cite{ref:grv_ho_pdfs} sets were chosen for the photon PDFs. 
Hadronisation corrections were obtained using the \texttt{Pythia}~\cite{ref:pythia} 
and \texttt{Herwig}~\cite{ref:herwig} Monte Carlo (MC) programs. For comparisons, samples of \texttt{Pythia} including 
multi-parton interactions~\cite{ref:pythia_mpi}, \texttt{Pythia-MI}, were used to estimate 
the contribution from non-perturbative effects not related to hadronisation. For all three jet algorithms introduced above, 
missing terms beyond NLO represent the dominant uncertainty of the predictions.

\section{Differential inclusive-jet cross sections}

Single- and double-differential inclusive-jet cross sections have been measured in the reaction $ep \to e + {\rm jet} + {\rm X}$ 
for $142 < W_{\gamma p} < 293$~GeV, where $W_{\gamma p}$ is the $\gamma p$ centre-of-mass energy, and $Q^{2} < 1$~GeV$^{2}$ with the ZEUS 
detector at HERA using an integrated luminosity of $300$~pb$^{-1}$. The cross sections include every jet with $\etjet > 17$~GeV and 
$-1 < \etajet < 2.5$~\cite{ref:zeus_incl_jets_php}. 

\begin{figure}[!h]
\begin{center}
\includegraphics[width=0.40\textwidth]{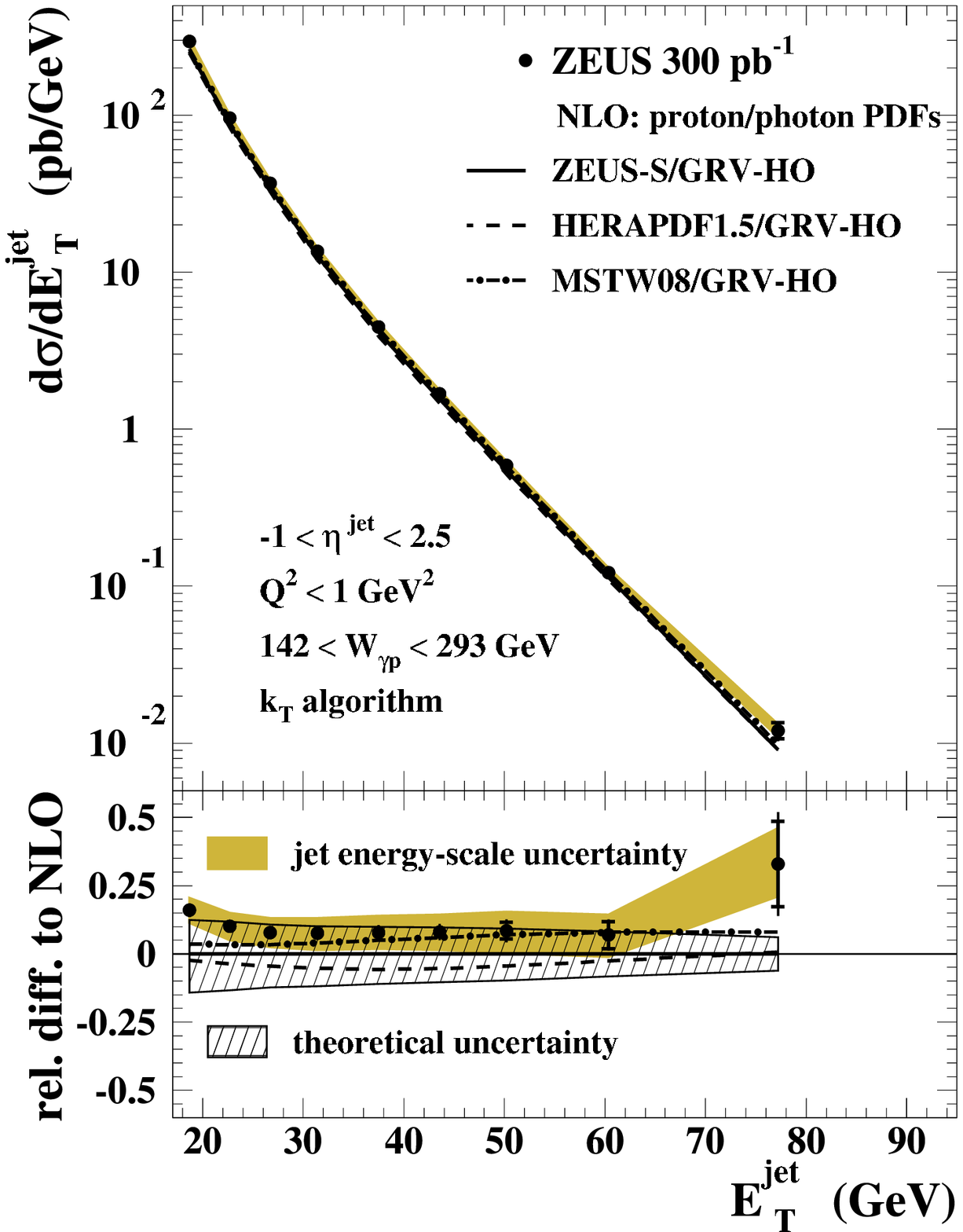} \hspace{2cm}
\includegraphics[width=0.40\textwidth]{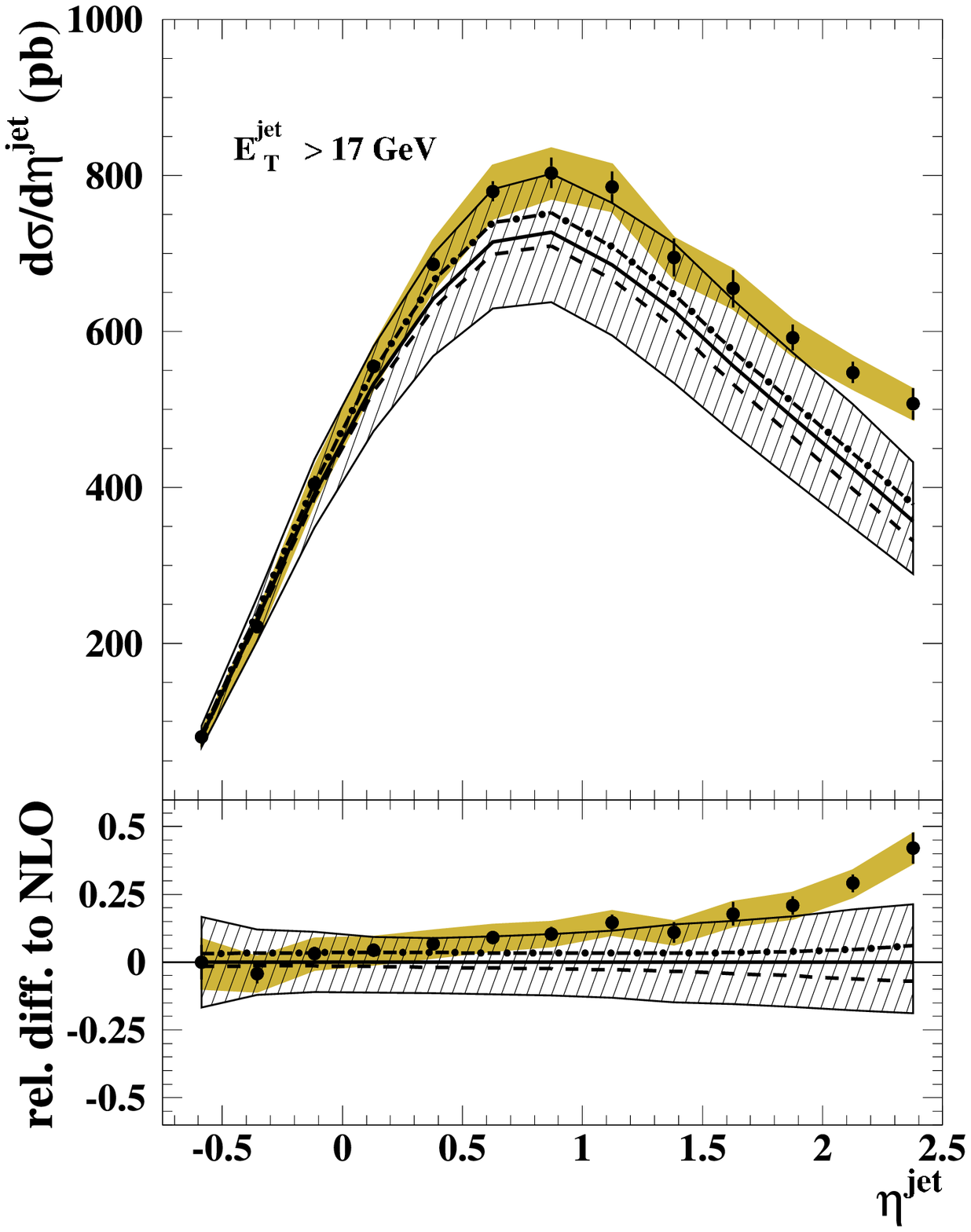}
\end{center}
\caption{Single-differential cross sections $d\sigma / d\etjet$ (left) and $d\sigma / d\etajet$ (right) based on the $k_{T}$ algorithm. The data are compared to NLO QCD predictions based on different proton PDFs.\label{fig:single_differential_cross_sections}}
\end{figure}

Single-differential cross sections based on the $k_{T}$ algorithm as functions of \etjet and \etajet 
are shown in Fig.~\ref{fig:single_differential_cross_sections}. The uncertainty on the jet energy scale of $\pm 1\%$ typically leads 
to a $\mp 5\%$ uncertainty on the measured cross sections which is fully correlated between measurements in different bins. 
At high \etjet this uncertainty increases to $\mp 10\%$. The measurements are well described by NLO QCD except 
for $\etajet > 2$. The disagreement in the forward region disappears if the kinematic region of 
the measurement is restricted to $\etjet > 21$~GeV.

Alternative NLO QCD predictions based on the HERAPDF1.5~\cite{ref:herapdf15_pdfs} and MSTW08~\cite{ref:mstw08_pdfs} proton PDFs 
instead of ZEUS-S are also shown in Fig.~\ref{fig:single_differential_cross_sections}. The predictions based on HERAPDF1.5 are 
lower than those based on ZEUS-S in most of the investigated phase-space region. Especially at large \etjet, the usage of MSTW08 
instead of ZEUS-S in the NLO QCD calculations leads to higher predictions. The high-precision measurements of inclusive-jet 
photoproduction have the potential to constrain the proton PDFs in future QCD fits.

In addition, inclusive-jet cross sections based on the $k_{T}$ algorithm were determined as functions 
of \etjet in different regions of \etajet. As observed for the single differential cross sections, the data 
are well described by NLO QCD except at $\etjet < 21$~GeV for $\etajet > 2$.

\section{Impact of multi-parton interactions}

\begin{figure}[!h]
\begin{center}
\includegraphics[width=0.40\textwidth]{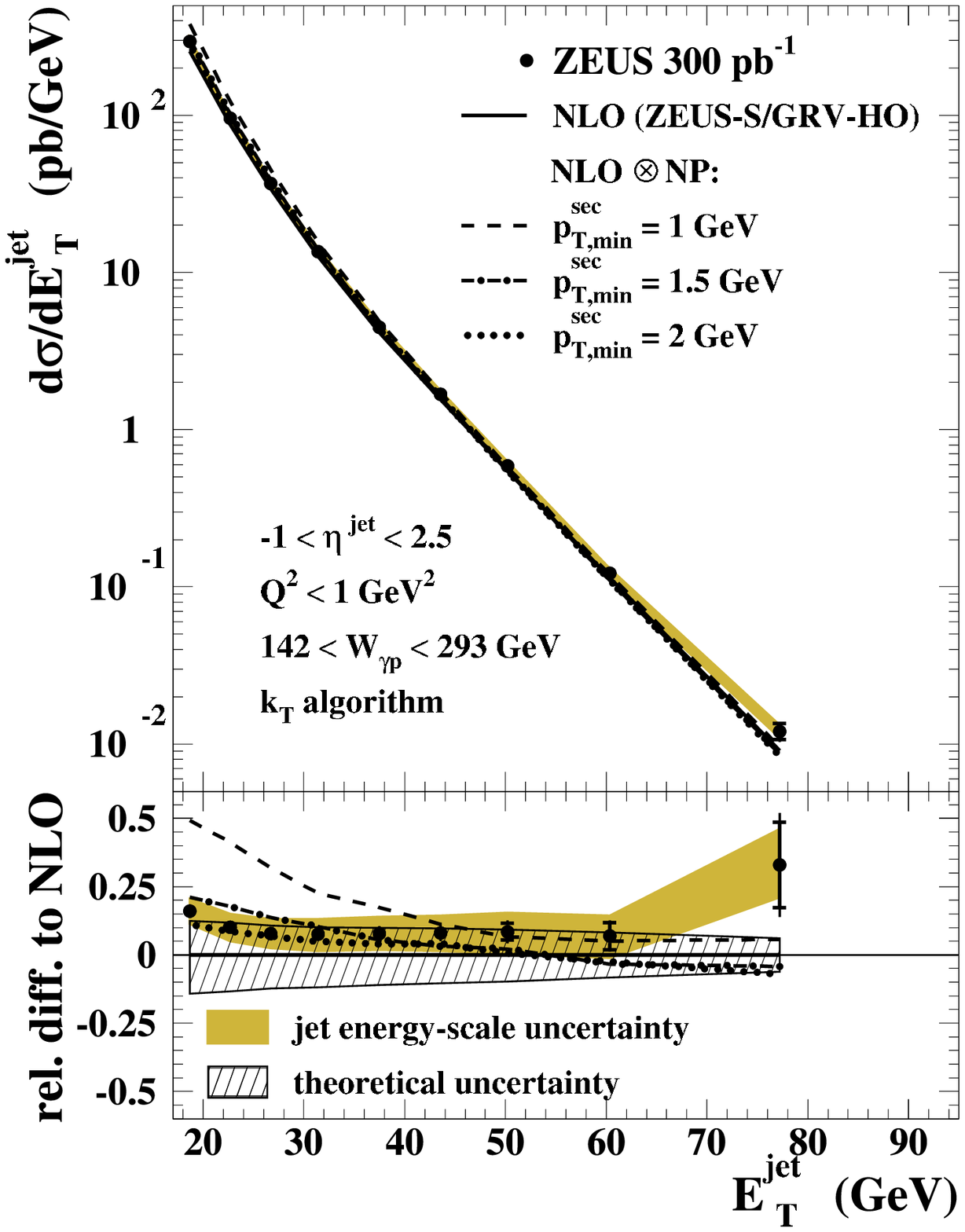} \hspace{2cm}
\includegraphics[width=0.40\textwidth]{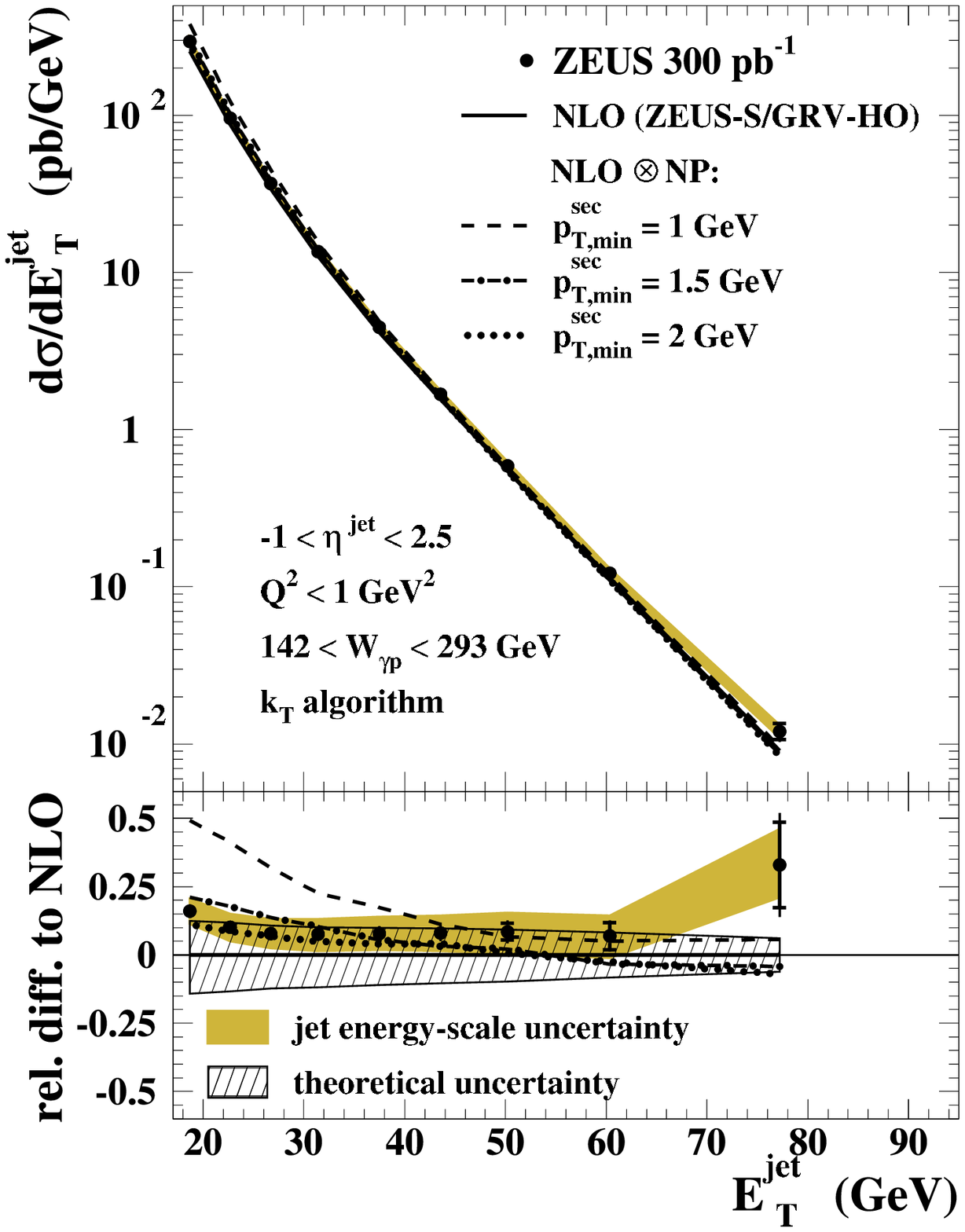}
\end{center}
\caption{Single-differential cross sections $d\sigma / d\etjet$ (left) and $d\sigma / d\etajet$ (right) based on the $k_{T}$ algorithm. The data are compared to NLO QCD predictions. For comparison, the NLO QCD calculations including an estimation of non-perturbative effects are shown in addition.\label{fig:single_differential_cross_sections_mpi}}
\end{figure}

The effect of multi-parton interactions is not included in the NLO QCD calculations described in Sec.~\ref{sec:introduction}. 
Instead, correction factors were obtained using \texttt{Pythia-MI} including multi-parton interactions with a minimum 
transverse momentum of the secondary scatter, $p_{T,{\rm min}}^{\rm sec}$, of 1, 1.5 and 2~GeV. Single-differential 
cross sections based on the $k_{T}$ algorithm as functions of \etjet and \etajet are compared to NLO QCD predictions where these 
correction factors have been applied are shown in Fig.~\ref{fig:single_differential_cross_sections_mpi}. The 
inclusion of multi-parton interactions increase the predictions at low \etjet and large \etajet. The best 
description of the data is observed for $p_{T,{\rm min}}^{\rm sec} = 1.5$~GeV. 

\section{Comparison of different jet algorithms}

Differential cross sections for inclusive-jet photoproduction as functions of \etjet and \etajet 
were measured for the $k_{T}$, anti-$k_{T}$ and SIScone jet algorithms. The hadronisation corrections are 
largest for the SIScone algorithm while similar corrections were found for the $k_{T}$ and anti-$k_{T}$ algorithms. 
As shown for the $k_{T}$ algorithm above, the measurements based on anti-$k_{T}$ and SIScone are well described 
by NLO QCD except at large \etajet.

\begin{figure}[!h]
\begin{center}
\includegraphics[height=4.8cm]{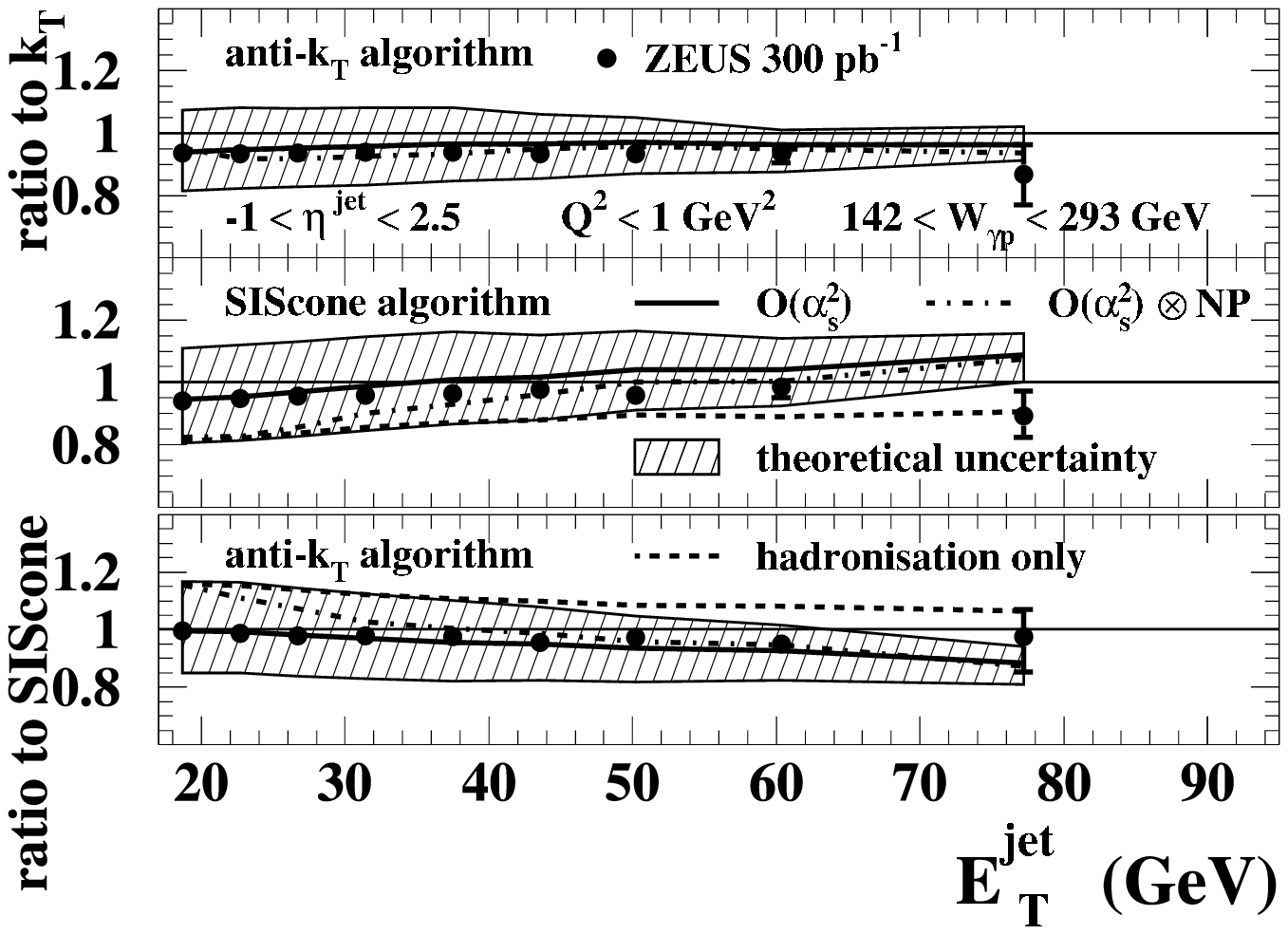} \hspace{1cm}
\includegraphics[height=4.8cm]{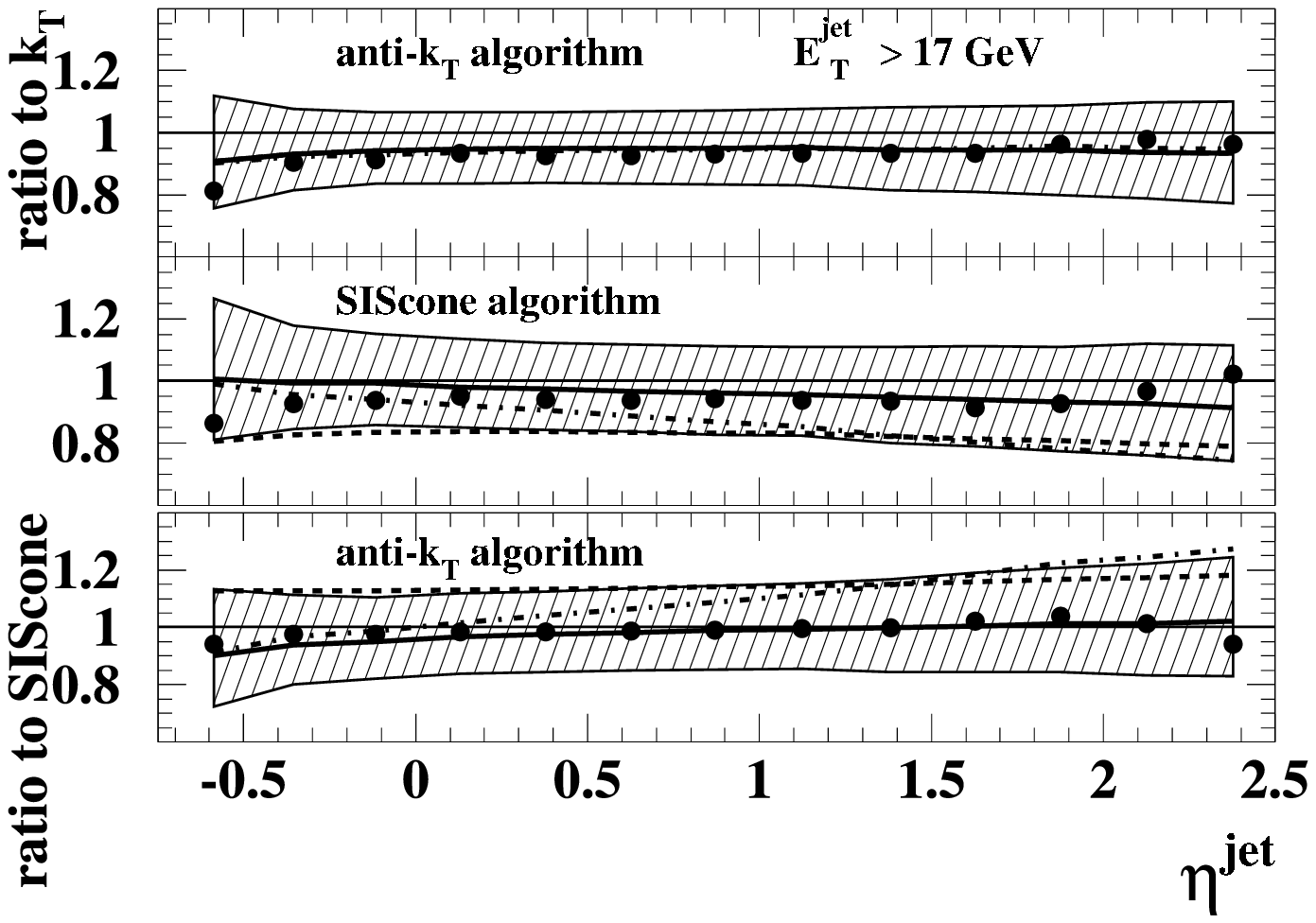}
\end{center}
\caption{The ratios of the measured cross sections anti-$k_{T}$/$k_{T}$, SIScone/$k_{T}$ and anti-$k_{T}$/SIScone as functions of \etjet (left) and \etajet (right).\label{fig:cross_sections_ratios}}
\end{figure}

To compare the different jet algorithms in detail, 
the rations of the measured cross sections anti-$k_{T}$/$k_{T}$, SIScone/$k_{T}$ and anti-$k_{T}$/SIScone were 
determined and are shown in Fig.~\ref{fig:cross_sections_ratios}. The cross sections for anti-$k_{T}$ have the same 
shape as those for $k_{T}$, but are about $6\%$ smaller. The measured cross sections based on SIScone have a slightly 
different shape than those based on $k_{T}$ or anti-$k_{T}$. The QCD calculations with up to three partons in the 
final state describe the measured ratios.

\section{Determination of $\alpha_{s}$ and its energy-scale dependence}

The measured single-differential cross sections $d\sigma / d\etjet$ for $21 < \etjet < 71$~GeV 
based on the $k_{T}$, anti-$k_{T}$ and SIScone jet algorithms were used to 
determine $\alphasmz$~\cite{ref:alphas_method}. Consistent results were obtained for all three 
jet algorithms:

\begin{eqnarray}
\alphasmz|_{k_{T}} &=& 0.1206\ ^{+0.0023}_{-0.0022}\ {\rm (exp.)}\ ^{+0.0042}_{-0.0035}\ {\rm (th.)},\nonumber\\
\alphasmz|_{{\rm anti-}k_{T}} &=& 0.1198\ ^{+0.0023}_{-0.0022}\ {\rm (exp.)}\ ^{+0.0041}_{-0.0034}\ {\rm (th.)},\nonumber\\
\alphasmz|_{\rm SIScone} &=& 0.1196\ ^{+0.0022}_{-0.0021}\ {\rm (exp.)}\ ^{+0.0046}_{-0.0043}\ {\rm (th.)}.\nonumber
\end{eqnarray}

\begin{figure}[!h]
\begin{center}
\includegraphics[width=0.60\textwidth]{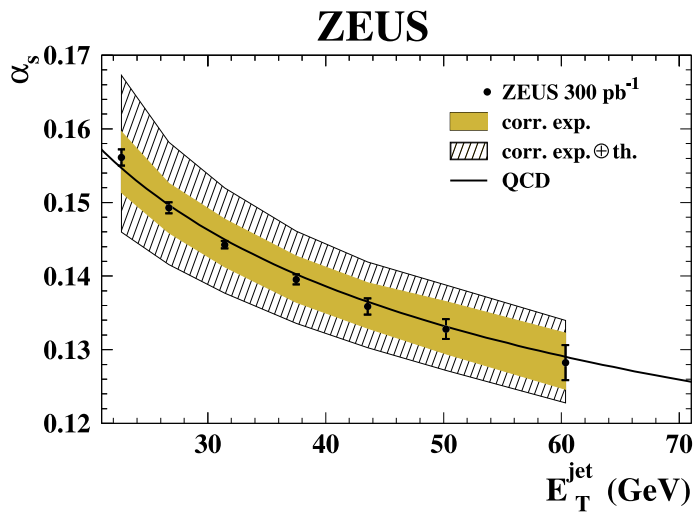}
\end{center}
\caption{$\alpha_{s}$ extracted at various $\langle \etjet \rangle$ values from the measured $d\sigma / d\etjet$ cross sections based on the $k_{T}$ algorithm.\label{fig:alphas_scale_dependence}}
\end{figure}

The results are in agreement with other determinations of \alphasmz~\cite{ref:zeus_incl_jets_php}. In addition, values of \alphas were extracted at 
the mean values, $\langle \etjet \rangle$, of the bins in \etjet without assuming the running of \alphas. The extracted 
values of \alphas as a function of \etjet are shown in Fig.~\ref{fig:alphas_scale_dependence}. This measurement confirms the 
running of \alphas over a wide \etjet range. The observed running is in good agreement with the two-loop QCD prediction.

\section{Summary and conclusions}

Inclusive-jet cross sections in photoproduction were measured using the ZEUS detector. The data are generally 
well described by NLO QCD predictions. The inclusion of multi-parton interactions improves the predictions at 
low \etjet and large \etajet. The presented measurements have the potential to improve the photon and proton PDFs in 
future QCD fits. The strong coupling constant was extracted at the $Z$ mass with competitive precision compared to other 
measurements and over a wide \etjet range.


\begin{thebibliography}{99}

\bibitem{ref:kt_algorithm} S.~Catani et al., \emph{Longitudinally-invariant $k_{\perp}$-clustering algorithms for hadron-hadron collisions}, Nucl. Phys. {\bf B 406}, 187 (1993).

\bibitem{ref:long_inv_incl_mode} S.D.~Ellis and D.E.~Soper, \emph{Successive combination jet algorithm for hadron collisions}, Phys. Rev. {\bf D 48}, 3160 (1993).

\bibitem{ref:anti_kt_algorithm} M.~Cacciari, G.P.~Salam and G.~Soyez, \emph{The anti-$k_{t}$ jet clustering algorithm}, JHEP {\bf 04}, 063 (2008).

\bibitem{ref:siscone_algorithm} G.P.~Salam and G.~Soyez, \emph{A practical seedless infrared-safe cone jet algorithm}, JHEP {\bf 05}, 086 (2007).

\bibitem{ref:nlo_predictions} M.~Klasen, T.~Kleinwort and G.~Kramer, \emph{Inclusive jet production in $\gamma p$ and $\gamma\gamma$ processes: direct and resolved photon cross sections in next-to-leading order QCD}, Eur. Phys. J. {\bf C 1}, 1 (1998).

\bibitem{ref:zeus_s_pdfs} S.~Chekanov et al., \emph{ZEUS next-to-leading-order QCD analysis of data on deep inelastic scattering}, Phys. Rev. {\bf D 67}, 012007 (2003).

\bibitem{ref:grv_ho_pdfs} M.~Gl\"{u}ck, E.~Reya, A.~Vogt, \emph{Parton structure of the photon beyond the leading order}, Phys. Rev. {\bf D 45}, 3986 (1992); \\
M.~Gl\"{u}ck, E.~Reya, A.~Vogt, \emph{Photonic parton distributions}, Phys. Rev. {\bf D 46}, 1973 (1992).

\bibitem{ref:pythia} T.~Sj\"{o}strand, \emph{High-energy-physics event generation with PYTHIA 5.7 and JETSET 7.4}, Comput. Phys. Comm. {\bf 82}, 74 (1994).

\bibitem{ref:herwig} G.~Marchesini et al., \emph{HERWIG 5.1 - a Monte Carlo event generator for simulating hadron emission reactions with interfering gluons}, Comput. Phys. Comm. {\bf 67}, 465 (1992); \\
G.~Corcella et al., \emph{HERWIG 6: an event generator for hadron emission reactions with interfering gluons (including supersymmetric processes)}, JHEP {\bf 01}, 010 (2001).

\bibitem{ref:pythia_mpi} T.~Sj\"{o}strand and M. van Zijl, \emph{A multiple-interaction model for the event structure in hadron collisions}, Phys. Rev. {\bf D 36}, 2019 (1987).

\bibitem{ref:zeus_incl_jets_php} H.~Abramowicz et al., \emph{Inclusive-jet photoproduction at HERA and determination of $\alpha_{s}$}, Nucl. Phys. {\bf B 864}, 1 (2012) [{\tt arXiv:1205.6153}].

\bibitem{ref:herapdf15_pdfs} F.D.~Aaron, \emph{Combined measurement and QCD analysis of the inclusive $e^{\pm}$ scattering cross sections at HERA}, JHEP {\bf 01}, 109 (2010).

\bibitem{ref:mstw08_pdfs} A.D.~Martin et al., \emph{Parton distributions for the LHC}, Eur. Phys. J. {\bf C 63}, 189 (2009).

\bibitem{ref:alphas_method} S.~Chekanov et al., \emph{Inclusive jet cross sections in the Breit frame in neutral current deep inelastic scattering at HERA and determination of \alphas}, Phys. Lett. {\bf B 547}, 164 (2002).

\end{thebibliography}
\end{document}